# Low-voltage Driving Phototransistor Based on Dye-sensitized Nanocrystalline Titanium Dioxide Thin Films

Xiaoqi Wang[a], Jia Xu and Chuanbing Cai

Physics Department, Shanghai University, Shanghai 200444, China

**Abstract:**

Photo-gated transistors based on dye-sensitized nanocrystalline titanium dioxide thin film are established. A transistor-like transport behavior characterized by the linear increase, saturated plateau, and breakdown-like increase in the voltage-current curve is achievable with a low driven bias for the present device. The response current exhibits a linear dependence on the intensity of gated light, and the measured maximum photosensitivity is approximately 0.1 A/W. The dynamic responses for various light frequencies and their dependences on the load resistances are investigated as well. The cut-off frequency of ~50 Hz is abstracted, indicating the potential application for economical and efficient light switch or optical communication unit. The *dc* photo-gated response is explained by the energy level diagram, and is numerically simulated by an equivalent circuit model, suggesting a clear correlation between photovoltaic and photoconductive behaviors as well as their optical responses.

**Keywords**：phototransistor; DSSC; frequency spectrum; equivalent electrical circuit;

[a]   Correspondence's E-mail: shawl.wang@gmail.com

## 1. Introduction

It is well known that the organic thin-film transistors (OTFTs) show their unique advantages over the conventional inorganic electronics, such as flexibility, light weight, and low cost etc, and thus they are attracting extensive interest recently[1-3]. During the past decade, there have been a lot of improvement on the performances, mechanism understanding, as well as the fabrication materials for the organic phototransistors (OPTs)[4-6]. For the operation with light, a back gate electrode is usually applied to incite the separation of the photo-excited electrons and holes in photoconducting unit[7]. This is called the charge separation process, which in most cases, results from the inner electric field, such as in inorganic or organic heterojunction photovoltaic cells[8]. In general, OPTs based on the conventional field-effect transistor structure have large operating voltages of 10 V or more[9-13]. This together with low field-effect drift mobility, usually results in too high power consumption and too low light sensitivity, to be practical for application. Further, unless the back-gated voltage is removed synchronously, OPTs have a too slow recombination rate to be operated in the dynamic optical communication[14-17]. Unlike the outside electric field in most OPTs mentioned above, a distinct charge separation mechanism induced by rapid electron injection and diffusion exists inside the Dye-sensitized Solar Cell (DSSC)[18-20], implying the possibility to gain emerging OPTs with the tailorable performances.

In this paper, we present a special phototransistors (PTs) consisting of two anti-series DSSCs. The clear dc and ac photo-gated response behavior are identified, and the energy level diagram is illustrated for the analysis of dc response mechanism, as well as **a** corresponding equivalent circuit model is briefly established for numerical simulation of the response behavior. The detailed calculation for the equivalent circuit model is stated in the Appendix.

## 2. Experimental details

A simple electric-burning method was used to separate the hard conducting layer of F-doped $SnO_2$ (FTO) into two parts, as shown in Fig.1 (a). This method employs one electrode connecting with the edge of FTO, and the other with a needle scratching along the middle of FTO. The distance between side electrode and needle is set to be 2~3 cm. By applying a sufficient voltage within 15~25 V, an approximate 100 μm insulating groove was drawn by the needle.

A 5×5 mm$^2$ monolayer of $TiO_2$ nanoparticles was prepared by screen printing on the etched glass. The porous anatase $TiO_2$ film is achieved after sintering process at 500 ℃,. Fig.1 (b) shows the configure of a transistor-like structure consisting of the counter electrode (CE) opposite to the source (S) and drain (D) electrodes, as well as the involved $TiO_2$ nanoparticles near the insulating regions[21]. The device cell was finally sealed by a 25 μm-thick plastic spacer together with the platinized CE after sensitizing for 20 h in the 0.3 mM solution with the dye of N719 and ethanol. The resultant device was filled with the electrolytes including 0.1 M LiI, 0.08 M $I_2$, 0.5 M 4-tert-butyl pyridine, and 0.6 M 2-Dimethyl-3-propylimidazolium iodide in acetonitrile. The voltage-current curves are measured by Keithley 2420 digital meters.

An AM1.5 light was provided by a commercial solar simulator (SAN-EI XES-151S) equipped with a 150W Xenon lamp. The waveform in the frequency measurement is recorded by an oscilloscope.

**3. Results and discussion**

*3.1 DC performance for the photo-electric response behavior*

Figure 2 illustrated phototransistor performance of the present device under different intensities of gated light. The light intensity is adjusted from dark to an AM1.5 sun of 100 mW/cm$^2$ by a series of neutral density filters. In case of illumination, the response current is characterized by three types of behaviors: a linear dependence on the low bias ($V_{DS}$), a gradual transition to the saturation at moderate $V_{DS}$, and a breakdown-like surge at high $V_{DS}$. The response currents increases with the light intensity enhanced and the maximum light sensitivity is achieved as 0.1 A/W. In the case of dark, the *V-I* curve shows a Schottky like behavior. These are actually of the typical performances of a phototransistor.

To summarize the relationship between current $I_{DS}$ and the photo-gated intensity, we illustrate the plot of $I_{DS}$ versus light intensity at various biases in Figure 3. It is interesting that whatever in the linear increasing region or in the saturation region, the current is almost proportional to the light intensity for each bias. The table inset in Fig. 3, shows the photocurrent/dark current ratios of photocurrents, $P=I_{DS}/I_{dark}$, for the studied phototransistor. Absolute values of *P* are subject to the applied light intensity and voltage. It is revealed that the *P* under the broadband illumination of 100

mW/cm$^2$, can reach the pronounced values as high as 3340 and 1690, at the low voltages of 0.1 V and 0.2 V, respectively. Of more importance is that such *P* is achieved at low voltages less than 0.3, and that the low power is applicable in a wide range of inciting light intensity including those as weak as 3.5 mW/cm$^2$. TABLE 1 **compares** the performance of DSSC based PTs with other OPTs.in the previous literature. Due to the difference in the mechanism of photo-electric response, DSSC based PTs show its merit in the low voltage operation. To our knowledge, OPTs operated by such extremely low voltages are hardly found before.

*3.2 AC performance for the photo-electric response behavior*

For the practical application, what is the most interesting is the dynamic property of this kind phototransistor operated by a fixed light signal. The measurement of frequency response in our experiment is schematically illustrated in Figure 4. A moderate bias $V_{DS}$ (~0.7 V) has been applied to make sure the device to operate in the saturation region. A square wave light signal is achieved by the AM1.5 light source and the chopper (SR 540). The waveform of response current is characterized by voltage ($V_R$) of a load resistance of 470 ohm, which is recorded by an oscilloscope. As the chopping frequency increasing, the output waveform becomes distorted gradually, from the square shape to triangle-like one, seen in Fig. 5 a, b, c, d. It appears to be related to the *RC* effect of DSSC-based transistor and load resistance. This can be understood if one considers that there is a capacitance effect in the DSSC cell as it works in a dynamic application[22]. The capacitance denotes the charging (discharging) process at the interface of TiO2/dye(electrolyte) with turning on (off) the light, i.e., as switched in illumination, a mount of electrons are injected into the $TiO_2$ and simultaneously a quantity of cations and "holes", such as $Li^+$ and $I_3^-$, are produced through absorption and redox reaction at the interface between the $TiO_2$ film and electrolyte.

To ascertain the RC effect, the normalized frequency spectrum for the studied device with various load resistances have been achieved, as indicated in Figure 6. As the load resistance becomes smaller, the response amplitude decays more slowly. Serially with a load $R$, the dynamic response of phototransistor is subjected to the $RC$ effect besides the inner recombination. For comparison with the traditional transistor, it is defined that the cut-off frequency is a frequency at which signal decreases down -3 dB. As shown, the DSSC based device may be limited for application in high frequency region when it is series connected with a load of large resistance, but it is still higher than the line frequency of 50 Hz ~ 60 Hz at a load of 1.3 Kohm, indicating the application for the present device with a high photoelectric transfer efficiency in daily life.

*3.3 Analysis for the energy band*

The transport mechanism for the present photo-gated transistor can be understood by the energy diagram, the detail discussion is stated in ref. 21. In the case of illumination, as the bias is 0 V, the photo-excited current (orange arrow) is counterbalanced to the dark current (black arrow) in which electrons transfers from TiO2 to electrolyte, as shown in Fig. 7(a); as a nonzero bias is applied, the device is then in a non-equilibrium state, and results in the response current we measured, shown in Fig. 7(b). With the bias $V_{DS}$ increasing, the TiO$_2$/dye levels on source ($L_S$) move upwards relative to redox level ($L_R$), while TiO$_2$/dye levels on drain ($L_D$) goes down. The former process will enhance the dark current around the $L_S$, while the latter will depress the dark current around the $L_D$ and maintaining the photo-excited current. At a high $V_{DS}$, however, $L_D$ moves down so deep that electrons immediately transfer from electrolyte to $L_D$, which resulting that response current $I_{DS}$ increases rapidly and perform a breakdown like behavior. Here, we noted that under illumination the device is in a critical equilibrium state achieved by two counterbalances, i.e. photo-excited current and dark current, any bias applied will broke the equilibrium state and induce the response current. This is different from the conventional OPTs where the response current is mainly subject to the carrier mobility and is required much larger driven bias.

*3.4 Proposal for the equivalent circuit model*

To incorporate with the analysis of energy band diagram, an equivalent circuit model consists of two anti-series connected DSSCs is established. For the equivalent circuit of DSSC, there are amounts of works have been done[23-25]. In the case of DC transport, DSSC can be generally treated as a parallel circuit consists of a source and a diode (D1). The source represents the constant process of photo-electric conversion and the diode depicts the recombination process between the TiO$_2$ particles and electrolyte, respectively, seen in the Figure 8. For our device, the constant current sources (PI) provide the photovoltaic currents $I_{phD}$ and $I_{phS}$ for solar cell on the drain and source electrode respectively, the diode (D1) represents the dark current inside each unit, an additional diode (D2) should be taken into account for the breakdown-like process where electrons intermediately transfers from electrolyte to TiO$_2$ within a larger bias region. The series resistance $R$ here is considered for the FTO. The interaction between the two anti-series connected DSSC units can be ignored because the influence is considered to be less[26]. In this article, the discussion is limited in the illumination and within a moderate bias region, there are only photoelectric conversion source and D1 will be taken into account. The complete calculation contained D2 and $R_S$ will be discussed in the appendix.

For each cell unit in Drain and Source respectively, the current density can be approximately written as follow:

$$j = j_e \exp(\frac{V_{cell}}{V_T}) - j_{ph} \dots\dots\dots\dots\dots\dots\dots\dots\dots\dots\dots\dots (1)$$

Where, $j_{ph}$ is the photocurrent density or is called as the short-circuit current density, $j_e$ the constant for exchanging current, $V_{cell}$ the voltage difference between the counter electrode and drain or source electrode, $V_T$ is a sample's characteristic parameter which is considered to be subjected to the technology of preparation and materials themselves. In generally it is written as $nk_BT/q$[22, 23], where $n$ are the ideal factor which is almost within 1~2, characterizing the interface of TiO2/electrolyte, and $k_B$, $T$, $q$ are the Boltzmann constant, the absolute temperature and the electron charge, respectively. To calculate the current, it requires the active area to be taken into account for each cell. Following the Kirchhoff's circuit laws, the relationship eq. (2) between current $I_{DS}$ and bias $V_{DS}$ can be derived:

$$I_{DS} = \frac{I_{phS}[\exp(\frac{V_{DS}}{V_T})-1]}{1+\frac{A_S}{A_D}\exp(\frac{V_{DS}}{V_T})} \quad \ldots\ldots\ldots\ldots (2);$$

$$I_{DS} = \frac{j_{ph}}{V_T(\frac{1}{A_S}+\frac{1}{A_D})}V_{DS} \quad \ldots\ldots\ldots\ldots\ldots (3);$$

$$I_{DS} = I_{phD} \quad \ldots\ldots\ldots\ldots\ldots\ldots\ldots\ldots (4);$$

Where, $A_D$ and $A_S$ are the active area for the TiO$_2$ film on drain (D) and source (S) electrode, respectively. While the bias is small enough, $\exp(V_{DS}/V_T)$ can be approximately reduced to $1+V_{DS}/V_T$ so that a linear relation of the current $I_{DS}$ with the external bias $V_{DS}$ is derived, as shown in eq. (3). The increasing slope is related to the photoelectric current density $j_{ph}$, the sample's characteristic parameter $V_T$ and the area of $A_S$ and $A_D$, suggesting that one can improve the performance of this kind of phototransistor through carefully choosing the preparation conditions and methods. For the specific condition and method, the maximum slope can be achieved if the equal area, i.e. $A_S=A_D$, is designed. While the bias $V_{DS}$ varies in a moderate region, $\exp(V_{DS}/V_T)$ is apparently larger than one so that $I_{DS}$ is transited into a constant value $I_{phD}$, which is only dependent on the illumination conditions, as shown in eq. (4). While $V_{DS}$ is large, however, the energy band of TiO$_2$ ($L_D$) decreases deep enough so that electrons will intermediately transfer from electrolyte to TiO$_2$. Therefore, to simulate the whole current-voltage curves one should take the D2 and RS into account, seen in appendix. Figure 9 shows a well-fitted *I-V* curve for the experimental data in the illumination and dark through using our equivalent circuit model. The parameters adopted in simulation are reasonable: such as $j_{ph}$=0.9×10$^{-2}$ and $V_{cell}$=0.64 are abstracted from the photovoltaic measurement. $A_S$=0.08 and $A_D$=0.17 are the fixed parameters for device preparation. $V_T$=0.043 can be derived from the fitness, by which the idea factor *n*=1.65 is deduced.

## 4. Conclusion

The present work shows a special phototransistors consisting of two anti-series DSSC. Unlike the solar cells based on the conventional *p-n* heterojunction, DSSC exhibits a distinct mechanism for carrier separation, mainly dominated by the diffusion after the fast electron injection from photo-exited dye to nanocrystalline $TiO_2$ film. This allows it possible to establish novel photoconductive devices. Under illumination, the transport response of the present device is characterized by typical phototransistor behavior, including the linear increase, saturated plateau, and breakdown-like increase in current-voltage curve. Moreover, the response current exhibits a linear dependence on the intensity of gated light. The dynamic response is characterized as well under the various resistance loads. Although the response decays in high frequency light signal, the cut-off (-3 dB) is still larger than the line frequency 50~60 Hz, indicating that the present photoconductive device is applicable as an economical and efficient light-operated switch or optical communication unit. The photo-gated response is explained by the energy level diagram, and is numerically simulated by an equivalent circuit model, suggesting a clear correlation between photovoltaic and photoconductive behaviors as well as their light responses.

**Acknowledgement**

The authors thank Dr. Songyuan Dai and Linhua Hu at the Key Laboratory of Novel Thin Film Solar Cell, Chinese Academy of Sciences, and Liyuan Han at the National Institute for Materials Science, Japan (NIMS, Japan) for their help and useful discussions. This work is partly sponsored by the Ministry of Science and Technology of China (973 Projects, No. 2011CBA00105, and 863 Projects, No. 2009AA03Z204), the Science and Technology Commission of Shanghai Municipality (No. 10dz1203500). The supports from the Open Project of State Key Laboratory of Functional Materials for Informatics (Chinese Academy of Sciences), and the Shanghai Leading Academic Discipline Project (No. S30105) are gratefully acknowledged as well.

**Appendices**

To derive the complete equation for the *V-I* relation, the immediately injection process (D2 in the equivalent circuit of Fig. 8) should be taken into account. In this appendix, the series resistance (RS) is also taken into account for realistic measurement situation. To derive the $I_{DS}$-$V_{DS}$ relation, the equivalent circuit is divided into three parts: (1) the photoelectric current source and D1 in the electrode S; (2) the photoelectric current source, D1 and D2 in the electrode D; (3) the series resistance.

The photoelectric current sources generate the current $j_{ph}A_S$ in the S and $j_{ph}A_D$ in the D, where, $A_D$ and $A_S$ are the area of TiO$_2$ film in the D and S, respectively. The recombination process (D1) occurs at the interface of TiO$_2$ and electrolyte can be represented by the formalism of a diode with the exchange current density $j_e$, i.e. $j_eA_D[\exp(V_D/V_T)-1]$ in the electrode D and $j_eA_S[\exp(V_S/V_T)-1]$ in the S, where $V_D$ and $V_S$ are the potential difference around the D1 in the D and S, respectively. The immediately injection process (D2) can be similarly written as $j_{\bar{e}}A_D[\exp(-V_D/V_T)-1]$ with the reverse exchange current density $j_{\bar{e}}$.

According to the Kirchhoff's theorem, one has the simultaneous equation concerning the voltage and current, which can be algebraically carried out, see the eq. (A.1- A.3).

$$I + j_{ph}A_S - j_eA_S[\exp(V_S/V_T) - 1] = 0 \quad\quad\quad (A.1)$$

$$I - j_{ph}A_D + j_eA_D[\exp(V_D/V_T) - 1] - j_{\bar{e}}A_D[\exp(-V_D/V_T) - 1] = 0 \quad (A.2)$$

$$V_S - V_D = V_{DS} - I_{DS}R \quad\quad\quad (A.3)$$

$$I_{DS} = -(j_{ph} + j_e)A_S + A_S \exp\left(\frac{V_{DS} - I_{DS}R}{V_T}\right) \frac{j_{const} + \Delta}{2[A_S \exp\left(\frac{V_{DS} - I_{DS}R}{V_T}\right) + A_D]} \quad\quad (A.4)$$

Where, $\Delta = sqrt\{j_{const}^2 + 4j_e j_{\bar{e}} A_D [A_S \exp\left(\frac{V_{DS} - I_{DS}R}{V_T}\right) + A_D]\}$

and $j_{const} = (j_{ph} + j_e)(A_D + A_S) - j_{\bar{e}}$.

Under illumination, the calculated $I_{DS}$ can be reduced to the eq. (2) with a small or a moderate bias, showing a linearly increasing and a saturation plateau and in a larger bias it will perform a breakdown like behavior. The expression (A. 4) is also suitable for the dark situation by setting the $j_{ph}$ to be zero, which perform a Schottky like behavior, seen in Fig. 9.

FIGURE CAPTIONS:

Fig. 1.

Schematic diagrams for the preparation and architecture of the present device cell: (a) etching method using a needle. (b) device structure consisting of the counter electrode (CE) opposite to the source (S) and drain (D).

Fig. 2

Current $I_{DS}$ vs. bias $V_{DS}$ for the present phototransistor with various intensity of the gated light.

Fig. 3

Current $I_{DS}$ vs. light intensity measured at three biases: 0.1, 0.2 and 0.3 V. The On/Off of the photocurrent ($I_{DS}/I_{dark}$) is concluded as a table presented in the inset.

Fig. 4

Schematics for the frequency response measurement for the present phototransistor.

Fig. 5

Waveform for the present phototransistor is recorded at various frequency of the light signal. (a) recorded at 10 Hz, (b) at 30 Hz, (c) at 60 Hz, (d) at 200 Hz.

Fig. 6

Frequency spectrum for the present phototransistor at various load resistors. The cut-off frequency is defined: upon the cut-off frequency, the response output decreases down -3 dB.

Fig. 7

Schematic diagram for the various potential levels involving in the present device cells, (a) equilibrium state $V_{DS}=0$; (b) nonequilibrium state $V_{DS}>0$. The solid lines denote the dominant direction of electron flowing, and the dash lines denote the potential direction of electron movement. The different current is represented by colors: orange denotes the photocurrent, black the dark current and green the intermediately injection current at a larger bias.

Fig. 8

Schematic diagram for the proposed equivalent circuit model.

Fig. 9

Simulation of the equivalent circuit model. The blue curve is iteratively calculated from the circuit model, and the hollow square and circle are abstracted from the experimental data in the condition of 100 mW/cm² and dark, respectively. Here we set the reverse exchange current density $j_{\bar{e}}=80*j_e$, and the series resistance $R=10$.

# TABLE 1

|  | Operation bias region | Light sensitivity | $I_{DS}/I_{dark}$ ratio |
|---|---|---|---|
| **DSSC based PTs** | ~ 1 V | 0.1 A/W @ 1.5AM | >849 @ 1.5AM |
| **Other OPTs** | > 10 V [9~13] | 82 ~ $10^3$ A/W [9, 12] | 800~$10^5$ [9, 12,15] |

**Figure 1 Wang et al**

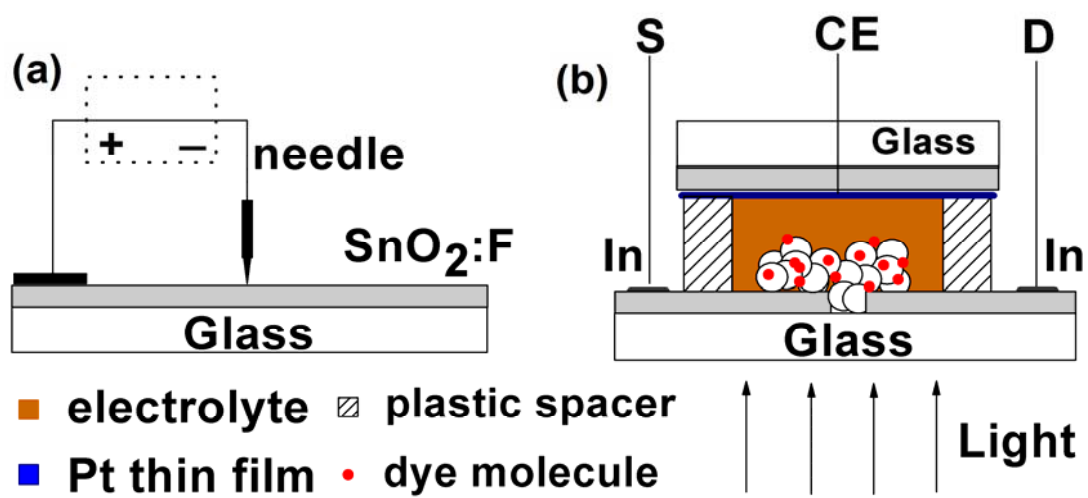

**Figure 2 Wang et al**

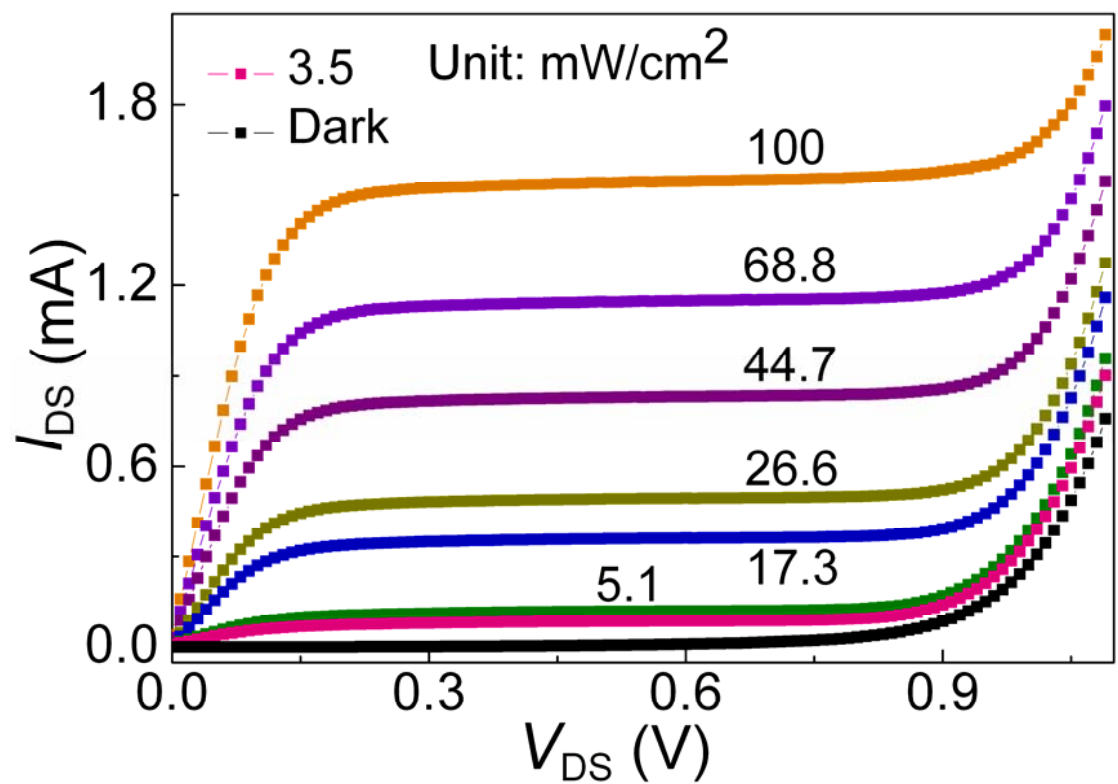

**Figure 3 Wang et al**

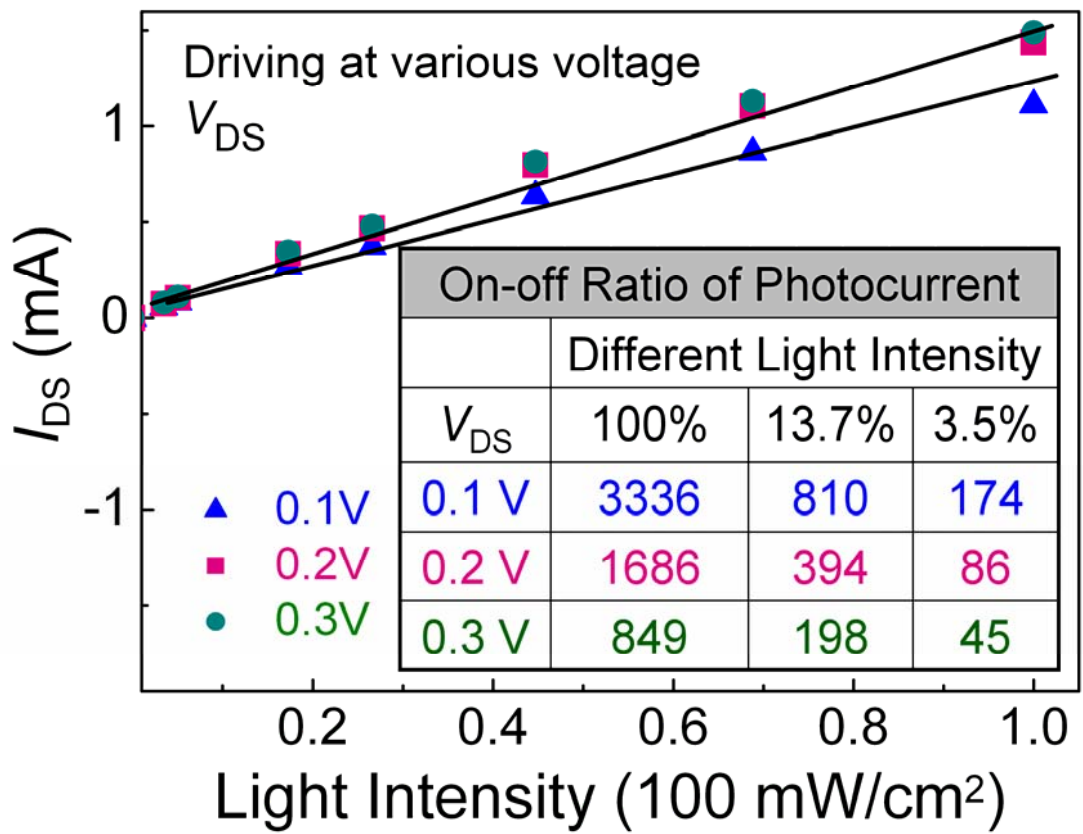

**Figure 4 Wang et al**

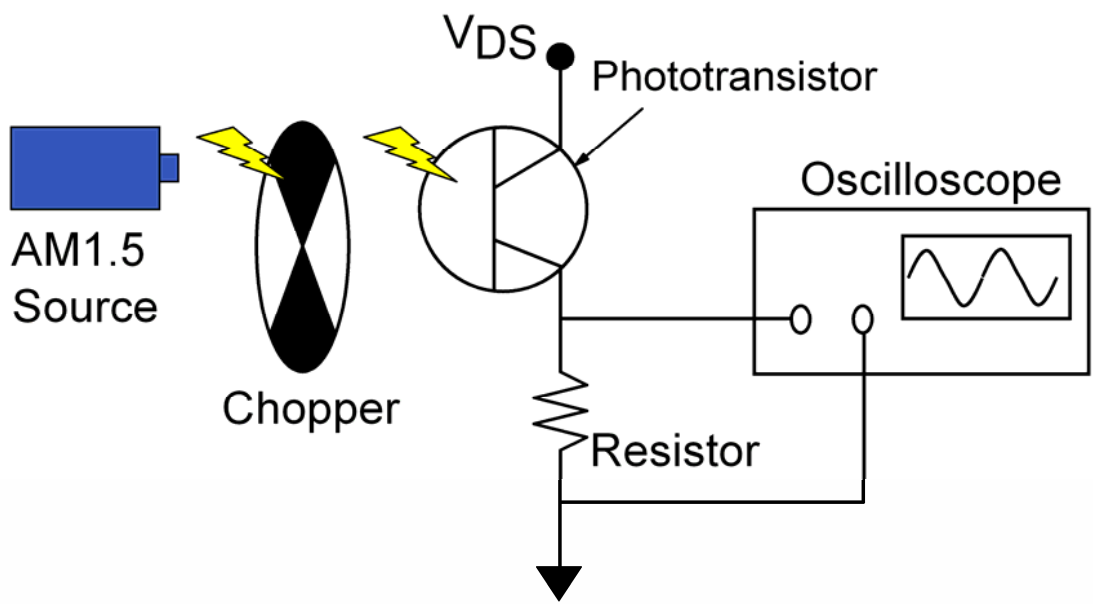

**Figure 5 Wang et al**

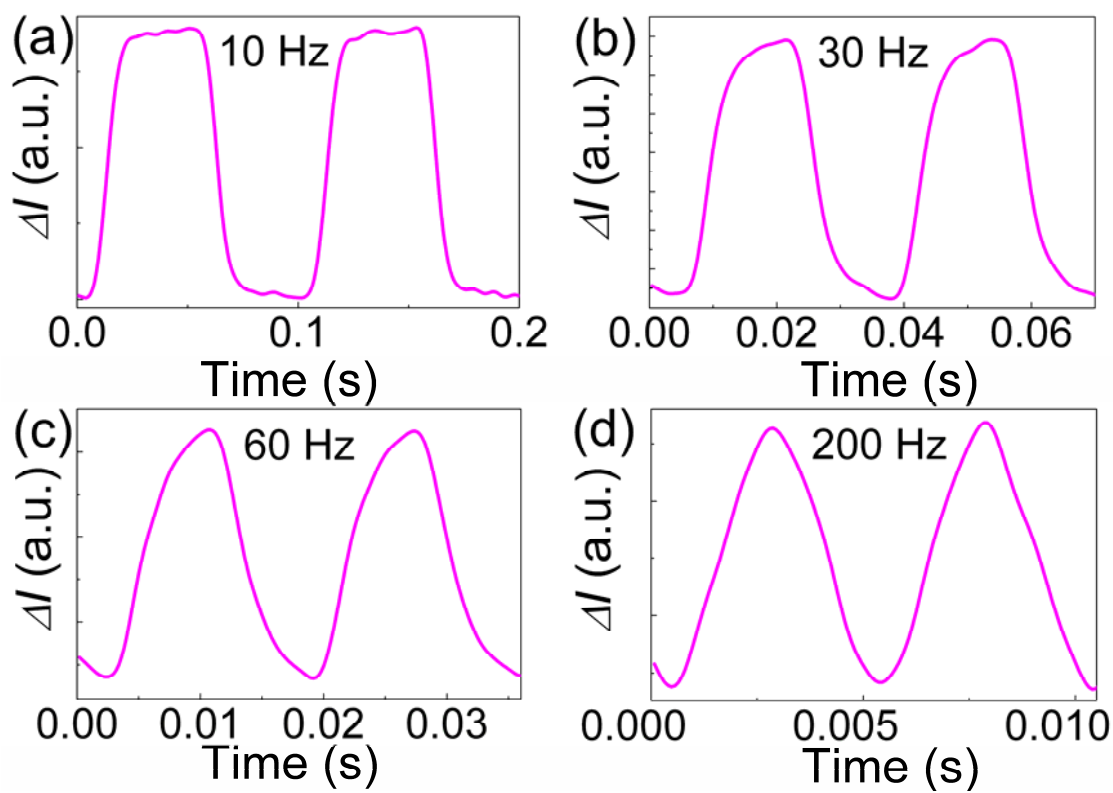

**Figure 6 Wang et al**

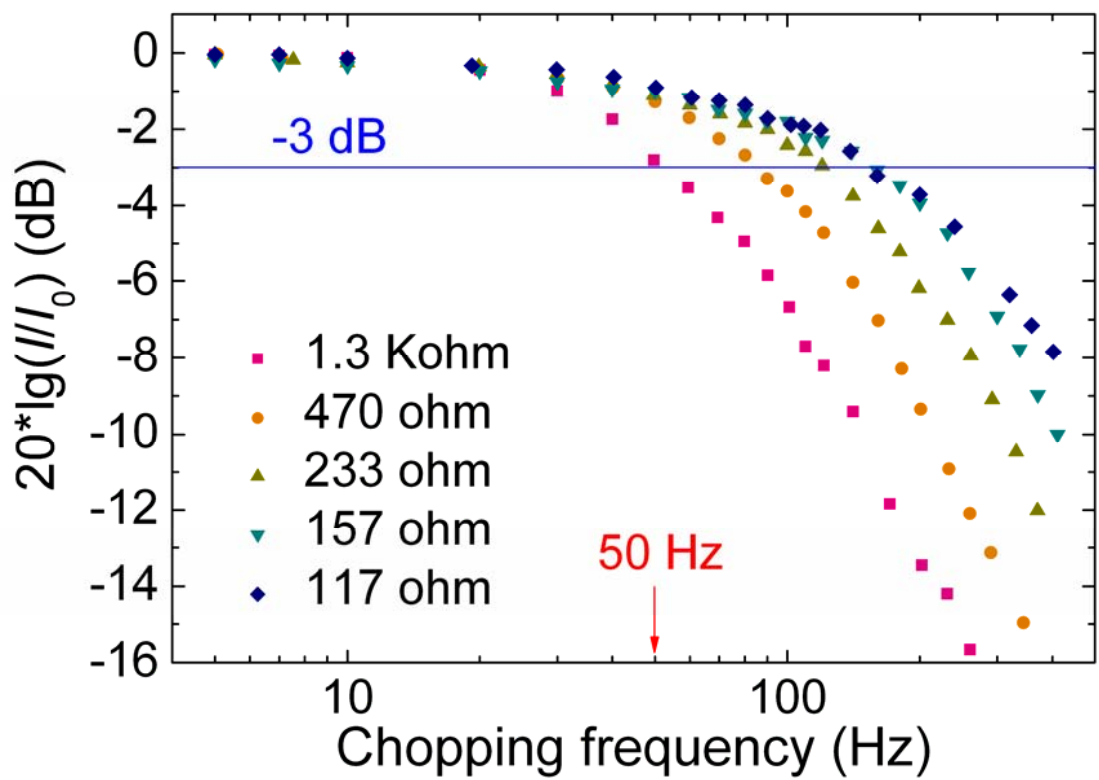

**Figure 7 Wang et al**

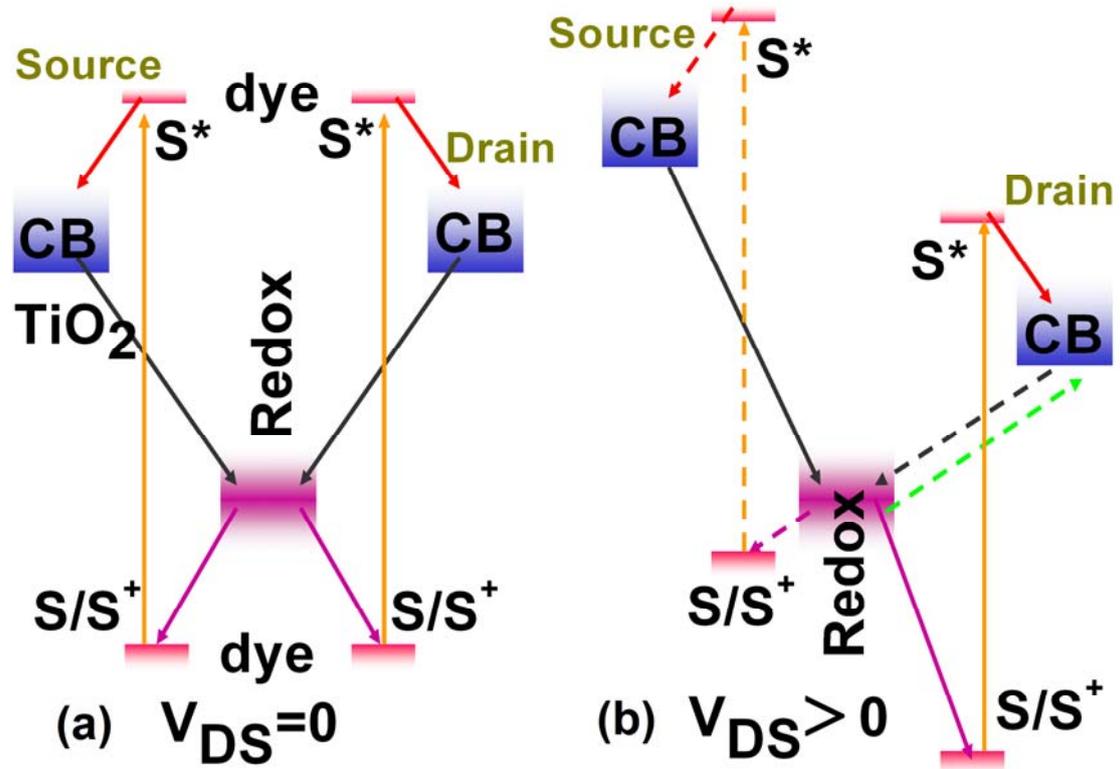

**Figure 8 Wang et al**

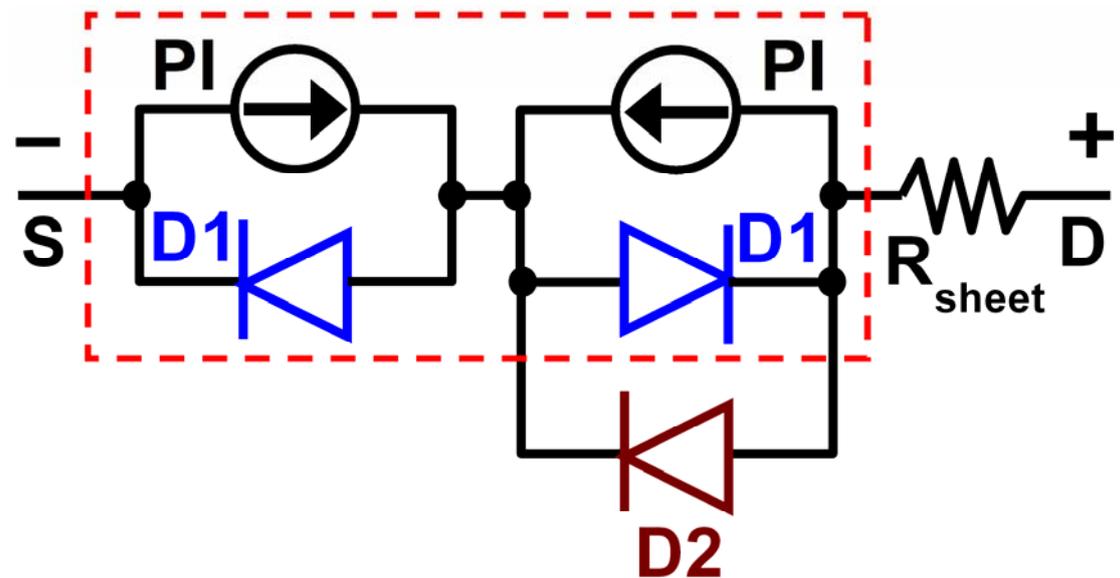

**Figure 9 Wang et al**

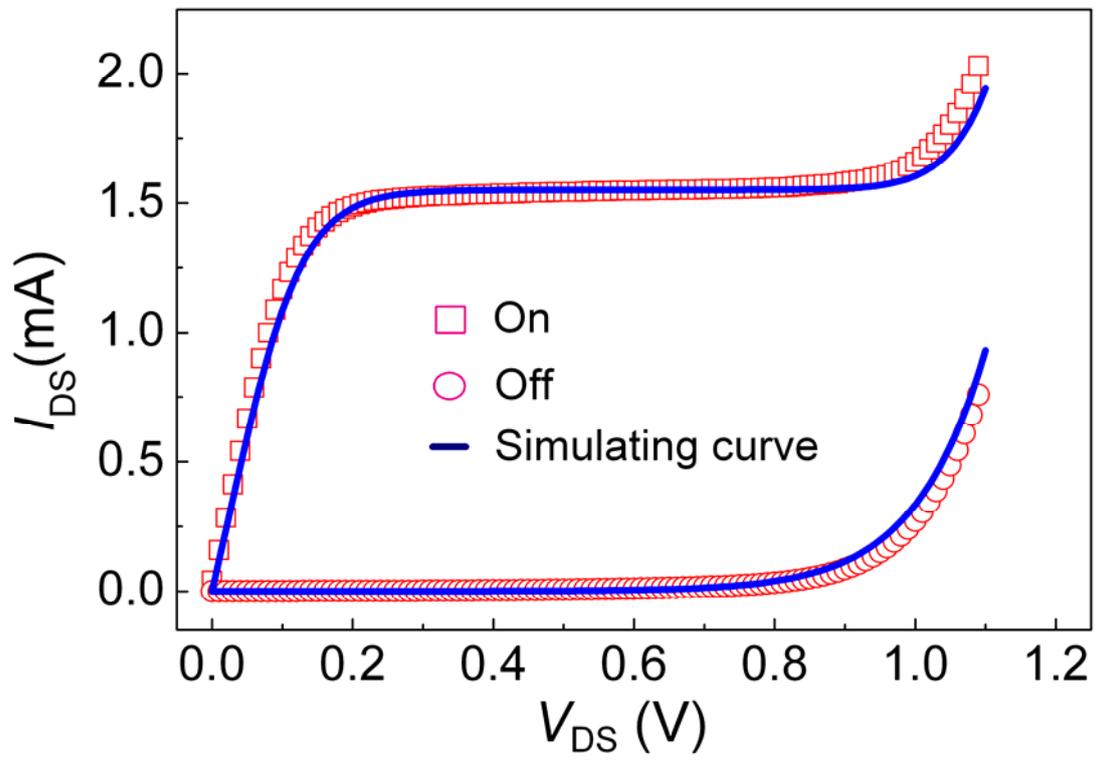

23